\pdfoutput=1
\documentclass[aps,nopreprint,twocolumn,prl,superscriptaddress,showpacs]{revtex4-1}
\usepackage{graphicx}
\usepackage{amsmath}
\usepackage{amsfonts}
\usepackage{hyperref}
\hypersetup{%
 pdftitle = {Ultralow frequency acoustic resonances and its potential for mitigating tsunami wave formation},
 pdfauthor = {H. Estrada, F. Meseguer},
 bookmarksopen = true,
 colorlinks  = true,
 citecolor   = black,
 linkcolor   = black,
 raiselinks  = true,
 urlcolor    = blue,
 pdfstartview= FitH,
}

\def\TT{{\rm T}}

\begin{document} 

\title{Ultralow frequency acoustic resonances and its potential for mitigating tsunami wave formation}

\author{H\'ector~Estrada}
\author{Francisco~Meseguer}
\email[Corresponding author: ]{fmese@fis.upv.es}
\affiliation{Centro de Tecnolog\'ias F\'isicas, Unidad Asociada ICMM- CSIC/UPV, Universidad Polit\'ecnica de Valencia, Av. de los Naranjos s/n. 46022 Valencia, Spain}
\affiliation{Instituto de Ciencia de Materiales de Madrid (CSIC), Cantoblanco, 28049 Madrid, Spain}
\date{\today}




\begin{abstract}
Bubbles display astonishing acoustical properties since they are able to absorb and scatter large amounts of energy coming from waves whose wavelengths are two orders of magnitude larger than the bubble size. Thus, as the interaction distance between bubbles is much larger than the bubble size, clouds of bubbles exhibit collective oscillations which can scatter acoustic waves three orders magnitude larger than the bubble size. Here we propose bubble based systems which resonate at frequencies that match the time scale relevant for seismogenic tsunami wave generation and may mitigate the devastating effects of tsunami waves. Based on a linear approximation, our na\"ive proposal may open new research paths towards the mitigation of tsunami waves generation.
\end{abstract}

\maketitle 

Tsunami waves are one of the most devastating natural events. Recent examples originated by submarine earthquakes having a magnitude $M_W>9.0$ have been covered by worldwide media showing the outcomes of such a natural disaster and the human and economic tragedy thereafter. Current measures and efforts are focused on forecast, early warning systems, and occasionally also on coastal tsunami run-up mitigation \cite{bernard2006}. Paradoxically, such destruction is generated by a very small amount of the strain energy released by the faulting, roughly speaking less than 1\% \cite{tang2012,lay2005}. Then, one may ask: Could it be possible to mitigate the generation of seismogenic tsunami waves?. 
Any thinkable mechanism that can be thought for this purpose acquires enormous proportions. Mitigation of coastal waves would imply huge barriers over extensive zones that are not affordable in the case of large tsunamis. It has been proposed the deployment solid periodic resonators arrays as a way to block water waves \cite{hu2011}. However, if applied to tsunami scales it concerns very large rigid scatterers hardly feasible for real situations. 

\begin{figure}[h]
 \begin{center}
  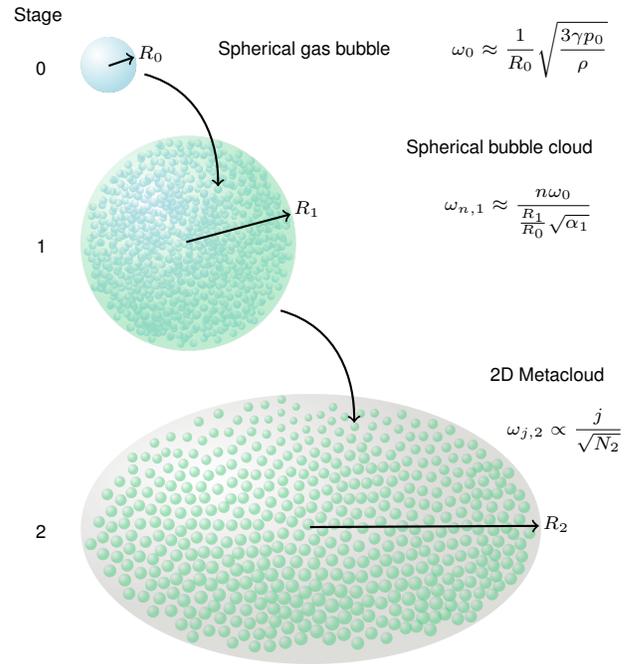
 \end{center}
\caption{\textbf{Diagram depicting the metacloud approach}. The building block (stage 0) consists of small air (gas) bubbles of radius $R_0$ which arranged form a spherical cloud (stage 1) of radius $R_1$ with a gas filling fraction $\alpha_1$. A number of $N_2$ clouds are then arranged within a disk (stage 2) of radius $R_2$ and thickness $2R_1$ with a cloud filling fraction $\alpha_2$ forming a metacloud. From the resonance formulae, $n,j=1,2,\dots$~. }
\label{fig:scheme}
\end{figure}

In a similar way to what happens in the sky, where kilometer size clouds composed by water droplets can be seen thank to light being scattered by tiny water droplets, we propose the use of air bubble metaclouds immersed into water to affect the generation of tsunamis (see Fig.~\ref{fig:scheme}). Thus, instead of dealing with a huge system composed of large parts, our proposal takes advantage of properly arranged small building blocks that could collectively resonate at a frequency scale a tsunami requires. 

\def\svgwidth{8.6cm}

Bubbles have unique properties due to the large mismatch between air and water mechanical compressibilities \cite{cavitation,acousticbubble}. A bubble (Fig.~\ref{fig:scheme} stage 0) having equilibrium radius $R_0$ displays a resonance \cite{minnaert1933} whose angular frequency is given by $\omega_0=R_0^{-1}\sqrt{3\gamma p_0/\rho}$, where $\rho$ corresponds to water density, $\gamma$ represents the specific heats ratio for air, and $p_0$ denotes the static pressure in water. Considering the sound speed in water as $c=1500$ m$/$s and atmospheric pressure, one can easily find that $\lambda_0/2R_0\approx 200$, which means that the wavelength in water at the bubble resonance is 200 times larger than its size. This deep subwavelength behavior makes bubbles a perfect building block for a metamaterial \cite{kafesaki2000}. In order to figure out the length scale of a bubble resonating with a tsunami wave, we should estimate the order of magnitude of the time scale of the seismic movement responsible of the tsunami formation. Long gravitational surface waves (tsunami waves in open ocean) in the linear regime propagate with a speed $c_g=\sqrt{gh}$ over the water layer of depth $h$ ($g=9.8$ m$/$s$^2$ is the acceleration due to gravity). Choosing the time interval that this wave requires to propagate a distance equal to $h$ as a target we obtain frequencies on the order of $\omega_\TT/2\pi=\sqrt{g/h}=0.075$~Hz for $h=5$~km. Using the resonance frequency of an hypothetical bubble tuned at $\omega_\TT$ we obtain a radius $R_\TT=\omega_\TT^{-1}\sqrt{3\gamma p_0/\rho}\approx43$~m. Although having a deep subwavelength resonance, this enormous bubble suffers from several practical drawbacks such as buoyancy that may be prevented by using huge gas balloons attached to the see bottom. However, the gradient pressure appearing at increasing sea depths should strongly affect the gas balloon sphericity, modifying its scattering properties.
 
Figure~\ref{fig:scheme} shows the strategy we have followed to reach a resonance frequency of $0.075$~Hz using a small bubble of few millimeters as building block (stage 0). In the next stage (stage 1) a bubble cloud is considered and eventually the metacloud stage (stage 2) is reached. Bubble clouds can be found in nature and have been studied as a source of noise in underwater environments \cite{prosperetti1988,lu1990}. In this context, collective modes of bubbles in clouds (Fig.~\ref{fig:scheme} stage 1) \cite{acousticbubble} have been observed due to the large acoustical scattering and absorption cross sections ($\sigma_s,\sigma_a$ respectively) of single bubbles allowing long range interactions across the cloud.

\begin{figure}[h]
 \begin{center}
  \includegraphics[width=8.6cm]{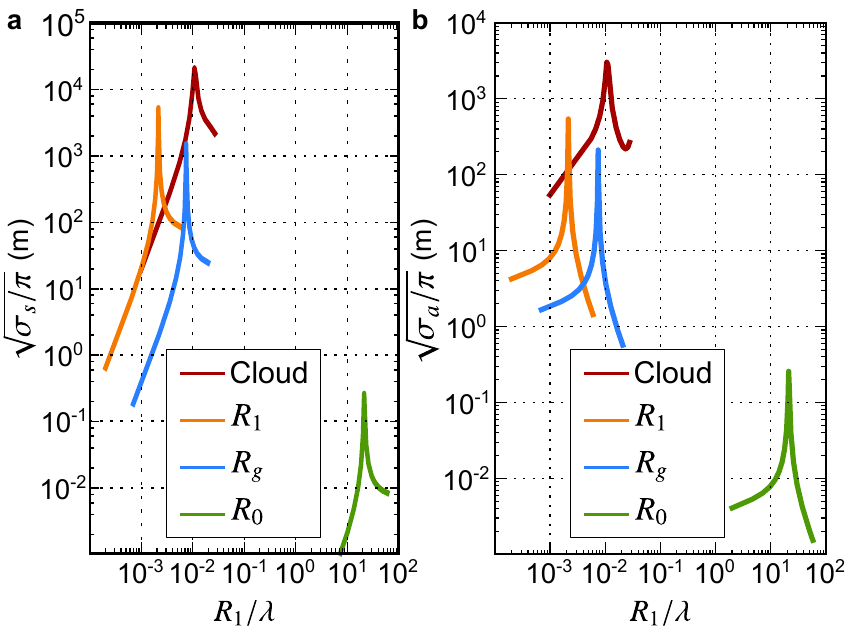}
 \end{center}
\caption{\textbf{Comparison between the interaction radii of a cloud and bubbles of different sizes}. Interaction radius $\sqrt{\sigma/\pi}$ in meters calculated from (\textbf{a}) scattering $\sigma_s$ and (\textbf{b}) absorption $\sigma_a$ cross sections for a cloud of radius $R_1=37$~m, a large bubble of the same radius $R_1$, a bubble of radius $R_g$ having the same volume of gas than the cloud, and a bubble of radius $R_0$ as a function of $R_1/\lambda$. The cloud has $R_1/R_0=10^4$ and $\alpha_1=0.024$.}
\label{fig:cross}
\end{figure}

\begin{figure}[h]
 \begin{center}
  \includegraphics[width=8.6cm]{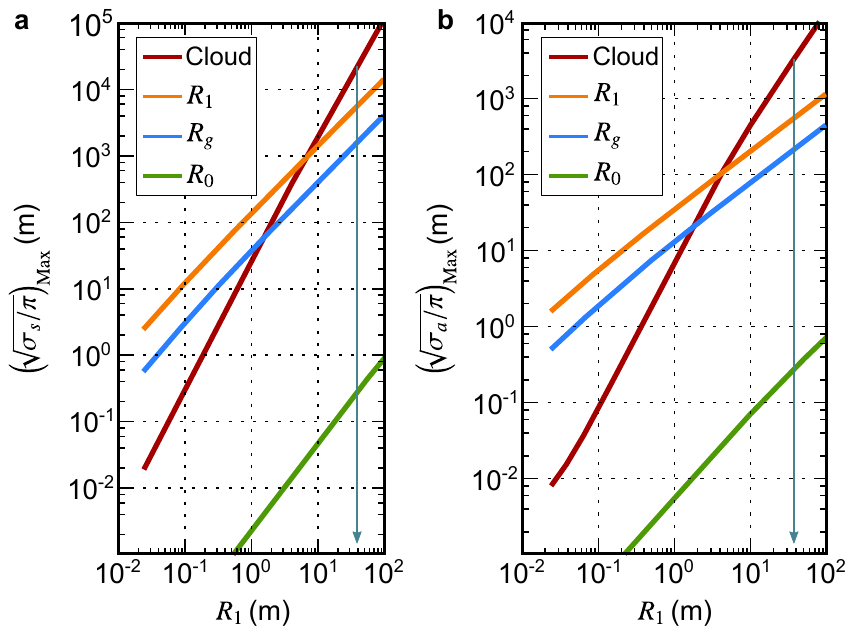}
 \end{center}
\caption{\textbf{Variation of the interaction radii as a function of the cloud radius}. Maximum acoustical interaction radius for (\textbf{a}) scattering and (\textbf{b}) absorption as a function of the cloud radius $R_1$ and compared to that of a bubble of the same size $R_1$, a bubble of radius $R_g$ having the same gas volume than the cloud, and a bubble of radius $R_0$, so that $R_1/R_0=10^4$ and $\alpha_1=0.024$. The vertical arrows indicate $R_1=37$~m, for which the interaction radius are plotted in Fig.~\ref{fig:cross}.}
\label{fig:rvar}
\end{figure}

Defining the acoustical interaction radius as $\sqrt{\sigma/\pi}$, we can compare between a spherical cloud and bubbles (or balloons) of different sizes as in Fig.~\ref{fig:cross}. Following \cite{prosperetti1977} and considering thermal effects and damping mechanisms, we calculate scattering and absorption interaction radius in the linear regime for time harmonic excitation, i.e. the bubble radius can be written as $R=R_0(1+\varphi\, {\rm exp}(i\omega t))$ when the complex oscillation amplitude $|\varphi|<<1$. Then, the bubbly water of the cloud \cite{dagostino1988}, can be treated as an effective homogeneous medium coupled to the surrounding liquid. A cloud of $R_1=37$~m formed by bubbles of $R_0=3.7$~mm with a gas filling fraction $\alpha_1=0.024$ clearly shows a collective resonant mode at $R_1/\lambda\approx0.01$ (Fig.~\ref{fig:cross} see labels). As it can be expected, the cloud interaction radius curves are orders of magnitude away from that of their building block. Also, the maximum interaction radius of the cloud overcome those provided by large bubbles having either the same radius $R_1$, or a radius $R_g=10.6$~m which contains the same volume of gas than the cloud. Although both large bubbles have lower resonant frequencies, the interaction radius is larger for the cloud, which depends strongly on the radius $R_1$. Taking $(\sqrt{\sigma/\pi})_{\rm Max}$ for different cloud radii $R_1$ while keeping $R_1/R_0=10^4$ and the same gas filling fraction for the cloud ($\alpha_1=0.024$), we have plotted in Fig.~\ref{fig:rvar} the maximum acoustical interaction radius of a cloud as a function of $R_1$. For the sake of comparison and in the same manner as in Fig.~\ref{fig:cross}, we also depict the cases of a single bubble (or balloon) of different sizes. It can be seen that above a critical radius ($R_1\approx10$~m) the cloud has higher interaction radius than the corresponding large bubbles. 

\begin{figure*}[ht]
 \begin{center}
  \includegraphics[width=\textwidth]{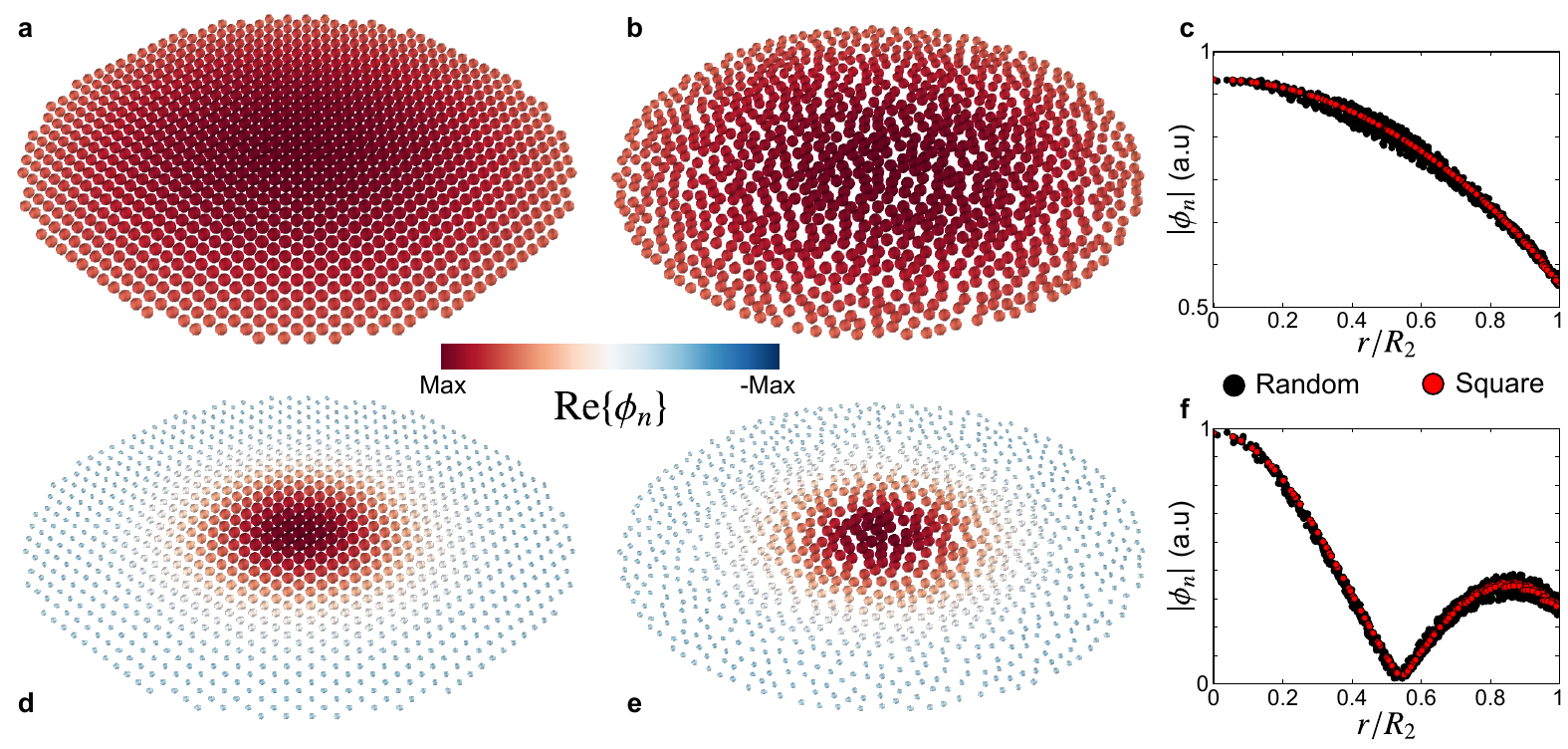}
 \end{center}
\caption{\textbf{Metacloud resonances}. Real part of the time harmonic component of the clouds pressure ${\rm Re}\{\phi_n\}$ (color scale and cloud size) in a metacloud for (\textbf{a}), (\textbf{d}) square and (\textbf{b}), (\textbf{e}) random metaclouds of radius $R_2=1.7$~km and a cloud filling fraction $\alpha_1=0.3$ and $N_1=1005$. (\textbf{c}), (\textbf{f}) $|\phi_n|$ in normalized units as a function of $r/R_2$ for both square and random arrays. The top row panels correspond to the first mode at $\omega_1/2\pi=0.075$~Hz and the bottom row ones to the second mode at $\omega_2/2\pi=0.15$~Hz. The clouds sizes are exaggerated to allow a proper visualization.} 
\label{fig:fields}
\end{figure*}

The acoustic results shown above allow us to consider low frequency modes of  clouds of bubbles as a single
bubble, i.e. the lowest resonant mode of a cloud  is similar to the fundamental breathing mode of a single bubble. Also the collective mode takes place when the wavelength in water is at least 50 times larger than the cloud radius  $R_1=37$~m \cite{dagostino1988}, which corresponds to a length around $1.85$~km. This wavelength value can further be enlarged to the tsunamis length scale by increasing the size of the bubble cloud. Alternatively, we can move further up to larger scales using the spherical cloud as a building block for a two dimensional (2D) metacloud (stage 2 in Fig.~\ref{fig:scheme}) and expect to see similar effects. In fact, using the multiple scattering method \footnote{As bubbles and clouds display deep subwavelength resonances, only monopolar behavior has been retained in the expansion.}\cite{morsefeshbach} we are able to observe the interaction between clouds within a metacloud, as depicted in Fig.~\ref{fig:fields}. Expressing the pressure at the $n$-cloud as $p_n=p_0+\phi_n\, {\rm exp}(i\omega t)$, then $Re\{\phi_n\}$ will give us information concerning the pressure at $t=0$ as depicted in Fig.~\ref{fig:fields}(\textbf{a}), (\textbf{b}), (\textbf{d}), and (\textbf{e}). Fig.~\ref{fig:fields}(\textbf{c}) and (\textbf{f}) show the time averaged pressure at the $n$-cloud $|\phi_n|$ as a function of its distance to the metacloud center normalized by the metacloud radius $r/R_2$. These two modes of the metacloud are obtained arranging $N_1=1005$ clouds to form either square or random two-dimensional arrays of radius $R_2=1.7$~km having a cloud filling fraction $\alpha_2=0.3$. 
Both modes have the same shape regardless the way the clouds are arranged (Fig.~\ref{fig:fields} (\textbf{a}) and (\textbf{b}); (\textbf{d}) and (\textbf{e})), which can be directly observed in Fig.~\ref{fig:fields} (\textbf{c}) and (\textbf{f}). In addition, there is a good qualitative agreement between the mode shape for spherical bubble clouds, for which $\phi(r) = \sin(kr)/kr$ as derived in \cite{dagostino1989} ($k$ is the wavenumber in the bubbly effective medium), and present results for a circular metacloud as shown in Fig.~\ref{fig:fields}(\textbf{c}) and (\textbf{f}). Thus, our results predict the existence of metacloud collective modes starting from large bubble clouds whose acoustical characteristics are comparable or even better than its large bubble counterparts. Consequently, a suitable spatial distribution of tiny bubbles having $3.7$~mm in size would be able to collectively oscillate at the low frequencies ($0.075$~Hz) of tsunamis.

In summary, within the linear approximation and using the acoustical approach, we predict the existence of collective oscillations of metaclouds. This phenomenon could be used to scale down the resonant frequency of a system able to target time scales which are characteristic in the generation of seismogenic tsunamis. Our proposal concerns a na\"ive approach as it avoids important ingredients such as bubble cavitation and nonlinear effects in clouds. An extended discussion on the role of incompressibility in the formation of seismogenic tsunami waves and metaclouds, the nonlinear behavior of bubbles and bubble clouds as well as some notes on the feasibility of the metacloud approach are given in the Appendix. We hope this work could open a new path towards tsunami generation mitigation, which, to the best of our knowledge, is not yet included in any tsunami-related agenda\cite{bernard2006}. Furthermore, our study could stimulate further investigation towards experimental demonstration of the collective oscillations of bubble metaclouds and the development of more realistic and quantitatively accurate theoretical models that may lead to a feasible strategy to mitigate tsunami generation. 

\begin{acknowledgments}
We thank E. Economou, J. Garcia de Abajo, and R. Alvarez for the critical reading of the manuscript. This work has been supported by the Spanish MICINN MAT2010-16879, Consolider CSD2007-00046 and Generalitat Valenciana PROMETEO 2010/043. F.M. conceived the idea of bubbles influencing tsunami wave generation. H.E. performed the calculations and developed the metacloud concept. F.M. and H.E. analyzed the data and wrote the paper.
\end{acknowledgments}

\appendix*
\section{Appendix}
\begin{figure}[h]
 \begin{center}
  \includegraphics[width=8.6cm]{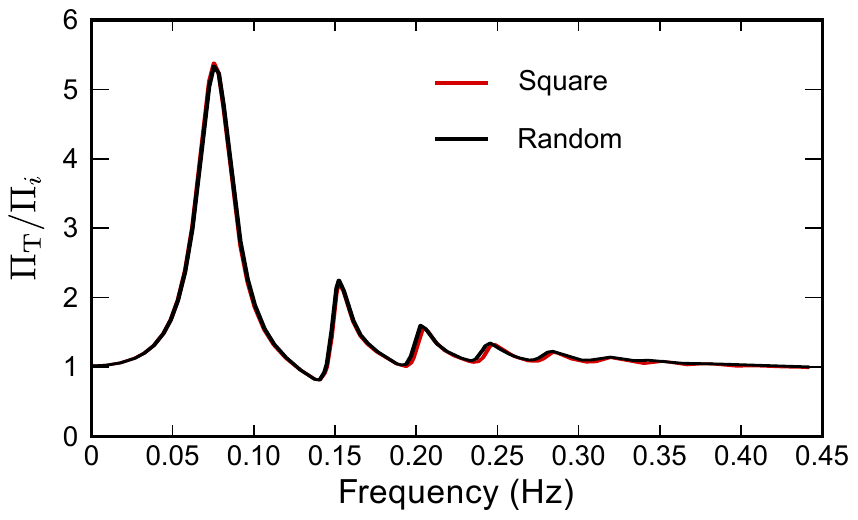} 
 \end{center}
\caption{Power balance of the two-dimensional metacloud ($R_2=1.7$~km, $R_1=37$~m, and $R_0=3.7$~mm) calculated using multiple scattering method under Born approximation.}
\label{fig:Born}
\end{figure}

\paragraph{Born approximation}
The effects of the bubble cloud on the acoustic wave propagation are difficult to deal with even within the linear regime. Born approximation, which is assumed in the multiple scattering formulation, limits the amplitude of the scattered wave to be negligible in comparison to the incident wave. In bubble clouds, however, this is not always the case and the model fails in giving quantitative predictions even at low filling fractions ($\alpha_1\sim0.01$) \cite{caflish1985b}. The power balance of the sum of the scattered and the absorbed power over the incident power $\Pi_{\mathrm T}/\Pi_{\mathrm i}$ deviates from unity as the Born approximation fails. However, the position of the peaks where the power balance is not preserved corresponds to the frequencies where collective modes appear. We have tested this behavior for clouds and expect the same to hold for metaclouds (see Fig.~\ref{fig:Born}). Thus, we obtain a qualitative estimation of metacloud collective modes depicted in Fig.~4 of the paper. More sophisticated methods have been developed for the one-dimensional case \cite{miksis1989} and to study the collective modes in microbubble clouds \cite{zeravcic2011} However, further work would be required to obtain quantitative predictions. 

\paragraph{Compressibility}
Tsunami waves generation by a moving sea-bottom is on its own a difficult problem. Considering the effect of compressibility in the generation of tsunami waves, the low resonant hydroacoustic mode of the water column at $h= \lambda/4$ can affect the transmission of the sea bottom displacement to the free water surface via nonlinear mechanism depending on the moving bottom velocity. In addition, the hydroacoustics mode would constitutes the only tsunami generation mechanism in the absence of sea bottom residual displacement \cite{tsunamis2009,nosov2007}. A distribution of bubbles (cloud or metacloud) located at a significant depth \footnote{When bubbles are at a depth of a few kilometers, due to the increased static pressure all the sizes discussed previously for the metacloud will increase in about one order of magnitude.} could affect this hydroacoustic modes. However, for slow sea-bottom velocities and large generation areas compared with the water depth, the water compressibility can be neglected and then the generation of tsunami waves can be understood within the incompressible water framework\cite{hammack1973,arcas2012}. One may think that as we obtained our results from the acoustic approximation and compressibility is a fundamental condition for acoustic waves to exist, metacloud and bubbles in general have no chance to affect tsunami generation when it is governed by incompressible mechanisms. However, due to \textbf{a)} the large differences between the compressibility of air and water and \textbf{b)} due to the deep subwavelength nature of the resonances studied, the incompressibility of water is also present in our proposal. The well known Rayleigh-Plesset equation, which is at the core of any study on air bubbles Moreover, in the spherical cloud model developed by d'Agostino and Brennen \cite{dagostino1988} incompressibility is recalled when deriving the scattering and absorption cross sections. We rederived their results in the linear regime including water compressibility in the boundary conditions and the only difference between their derivation and ours lies on  $(i2\pi R_1/\lambda +1)$ factors. Provided that $R_1/\lambda<<1$ (deep subwavelength regime) the incompressible approximation dominates. In simple words, as the wavelength is so large compared to the cloud (metacloud), the bubbles (clouds) only feel hydrostatic pressure across the whole ensemble. Thus, incompressible flow is not an issue that would turn off the collective metacloud resonance, although further research is certainly needed to know how the tsunami generation would be affected by these resonances.  
                                                                                                                                                                                                                                                                                                                                                                                                                                                                                                                                                                                                                                                                                                                                                                                                                                                                                                                                                                                                                                                                                                                                                                                                                                                                                                                                                                                                                                                                                                                                                                              \paragraph{Nonlinearity}                                                                                                                                                                                                                                                                                                                                                                                                                                                                                                                                                                                                                                                                                                                                                                                                                                                                                                                                                                                                                                                                                                                                                                                                                                                                                                                                                                                                                                                                                                                                                                It is well known that even for small amplitude, bubbles can display a rich nonlinear behavior \cite{acousticbubble}. Cavitating clouds have also being studied mainly in the context of hydrodynamic systems \cite{cavitation,acousticbubble}. How nonlinearities would affect the collective oscillations of the metacloud and whether these nonlinearities would play or not a role in tsunami generation mitigation are challenging questions that should be addressed in the future.

\paragraph{Feasibility}
There are several factors which come into play concerning the feasibility of the metacloud approach for tsunami generation mitigation. If small bubbles were to be chosen, polydispersity should not be an issue since the lowest order modes are only slightly affected \cite{zeravcic2011}. The generation of monodisperse microbubbles has been successfully achieved in laboratory environments \cite{ganan2001}, however, another source of polydispersity will arise if the small bubbles are supposed to form clouds having tens of meters given by the static pressure difference along the cloud. Buoyancy and interaction of the clouds with ocean currents should also be considered. If instead of small bubbles, gas filled balloons $\sim1$~m were to be chosen as the building block, buoyancy could be countered by means of ballast. In this case, the sphericity of the balloon would be compromised and its effect together with the effect of the covering membrane must be taken into account. Whether the cloud/metacloud should lie near to the sea surface or at a certain depth must be considered at the light of its influence on the tsunami generation. Although our proposal considers a certain degree of randomness within the metacloud, provided that a certain global filling fraction $\alpha_2$ is reached and the clouds are at a minimum distance of each other (smaller than the interaction radius), in a real deployment the appropriate geometry might be an stochastic fractal one.  


\begin{thebibliography}{10}%
\makeatletter
\providecommand \@ifxundefined [1]{%
 \ifx #1\undefined \expandafter \@firstoftwo
 \else \expandafter \@secondoftwo
\fi
}%
\providecommand \@ifnum [1]{%
 \ifnum #1\expandafter \@firstoftwo
 \else \expandafter \@secondoftwo
\fi
}%
\providecommand \enquote [1]{``#1''}%
\providecommand \bibnamefont  [1]{#1}%
\providecommand \bibfnamefont [1]{#1}%
\providecommand \citenamefont [1]{#1}%
\providecommand\href[0]{\@sanitize\@href}%
\providecommand\@href[1]{\endgroup\@@startlink{#1}\endgroup\@@href}%
\providecommand\@@href[1]{#1\@@endlink}%
\providecommand \@sanitize [0]{\begingroup\catcode`\&12\catcode`\#12\relax}%
\@ifxundefined \pdfoutput {\@firstoftwo}{%
 \@ifnum{\z@=\pdfoutput}{\@firstoftwo}{\@secondoftwo}%
}{%
 \providecommand\@@startlink[1]{\leavevmode}%
 \providecommand\@@endlink[0]{}%
}{%
 \providecommand\@@startlink[1]{%
  \leavevmode
  \pdfstartlink
   attr{/Border[0 0 1 ]/H/I/C[0 1 1]}%
   user{/Subtype/Link/A<</Type/Action/S/URI/URI(#1)>>}%
  \relax
 }%
 \providecommand\@@endlink[0]{\pdfendlink}%
}%
\providecommand \url  [0]{\begingroup\@sanitize \@url }%
\providecommand \@url [1]{\endgroup\@href {#1}{\urlprefix}}%
\providecommand \urlprefix [0]{URL }%
\providecommand \Eprint[0]{\href }%
\@ifxundefined \urlstyle {%
  \providecommand \doi [1]{doi:\discretionary{}{}{}#1}%
}{%
  \providecommand \doi [0]{doi:\discretionary{}{}{}\begingroup
  \urlstyle{rm}\Url }%
}%
\providecommand \doibase [0]{http://dx.doi.org/}%
\providecommand \Doi[1]{\href{\doibase#1}}%
\providecommand \bibAnnote [3]{%
  \BibitemShut{#1}%
  \begin{quotation}\noindent
    \textsc{Key:}\ #2\\\textsc{Annotation:}\ #3%
  \end{quotation}%
}%
\providecommand \bibAnnoteFile [2]{%
  \IfFileExists{#2}{\bibAnnote {#1} {#2} {\input{#2}}}{}%
}%
\providecommand \typeout [0]{\immediate \write \m@ne }%
\providecommand \selectlanguage [0]{\@gobble}%
\providecommand \bibinfo [0]{\@secondoftwo}%
\providecommand \bibfield [0]{\@secondoftwo}%
\providecommand \translation [1]{[#1]}%
\providecommand \BibitemOpen[0]{}%
\providecommand \bibitemStop [0]{}%
\providecommand \bibitemNoStop [0]{.\EOS\space}%
\providecommand \EOS [0]{\spacefactor3000\relax}%
\providecommand \BibitemShut [1]{\csname bibitem#1\endcsname}%
\bibitem{bernard2006}%
  \BibitemOpen
  \bibfield{author}{%
  \bibinfo {author} {\bibfnamefont{E.}~\bibnamefont{Bernard}}, \bibinfo
  {author} {\bibfnamefont{H.}~\bibnamefont{Mofjeld}}, \bibinfo {author}
  {\bibfnamefont{V.}~\bibnamefont{Titov}}, \bibinfo {author}
  {\bibfnamefont{C.}~\bibnamefont{Synolakis}},\ and\ \bibinfo {author}
  {\bibfnamefont{F.}~\bibnamefont{González}},\ }%
  \bibfield{journal}{%
  \Doi{10.1098/rsta.2006.1809}{\bibinfo {journal} {Phil. Trans. R. Soc. A}}\ }%
  \textbf{\bibinfo {volume} {364}},\ \bibinfo {pages} {1989} (\bibinfo {year}
  {2006}),\
  \bibAnnoteFile{NoStop}{bernard2006}%
\bibitem{tang2012}%
  \BibitemOpen
  \bibfield{author}{%
  \bibinfo {author} {\bibfnamefont{L.}~\bibnamefont{Tang}}, \bibinfo {author}
  {\bibfnamefont{V.~V.}\ \bibnamefont{Titov}}, \bibinfo {author}
  {\bibfnamefont{E.~N.}\ \bibnamefont{Bernard}}, \bibinfo {author}
  {\bibfnamefont{Y.}~\bibnamefont{Wei}}, \bibinfo {author}
  {\bibfnamefont{C.~D.}\ \bibnamefont{Chamberlin}}, \bibinfo {author}
  {\bibfnamefont{J.~C.}\ \bibnamefont{Newman}}, \bibinfo {author}
  {\bibfnamefont{H.~O.}\ \bibnamefont{Mofjeld}}, \bibinfo {author}
  {\bibfnamefont{D.}~\bibnamefont{Arcas}}, \bibinfo {author}
  {\bibfnamefont{M.~C.}\ \bibnamefont{Eble}}, \bibinfo {author}
  {\bibfnamefont{C.}~\bibnamefont{Moore}}, \bibinfo {author}
  {\bibfnamefont{B.}~\bibnamefont{Uslu}}, \bibinfo {author}
  {\bibfnamefont{C.}~\bibnamefont{Pells}}, \bibinfo {author}
  {\bibfnamefont{M.}~\bibnamefont{Spillane}}, \bibinfo {author}
  {\bibfnamefont{L.}~\bibnamefont{Wright}},\ and\ \bibinfo {author}
  {\bibfnamefont{E.}~\bibnamefont{Gica}},\ }%
  \bibfield{journal}{%
  \Doi{10.1029/2011JC007635}{\bibinfo {journal} {J. Geophys. Res.}\ }}%
  \textbf{\bibinfo {volume} {117}},\ \bibinfo {pages} {C08008} (\bibinfo {year} {2012}),\
  \bibAnnoteFile{NoStop}{tang2012}%
\bibitem{lay2005}%
  \BibitemOpen
  \bibfield{author}{%
  \bibinfo {author} {\bibfnamefont{T.}~\bibnamefont{Lay}}, \bibinfo {author}
  {\bibfnamefont{H.}~\bibnamefont{Kanamori}}, \bibinfo {author}
  {\bibfnamefont{C.~J.}\ \bibnamefont{Ammon}}, \bibinfo {author}
  {\bibfnamefont{M.}~\bibnamefont{Nettles}}, \bibinfo {author}
  {\bibfnamefont{S.~N.}\ \bibnamefont{Ward}}, \bibinfo {author}
  {\bibfnamefont{R.~C.}\ \bibnamefont{Aster}}, \bibinfo {author}
  {\bibfnamefont{S.~L.}\ \bibnamefont{Beck}}, \bibinfo {author}
  {\bibfnamefont{S.~L.}\ \bibnamefont{Bilek}}, \bibinfo {author}
  {\bibfnamefont{M.~R.}\ \bibnamefont{Brudzinski}}, \bibinfo {author}
  {\bibfnamefont{R.}~\bibnamefont{Butler}}, \bibinfo {author}
  {\bibfnamefont{H.~R.}\ \bibnamefont{DeShon}}, \bibinfo {author}
  {\bibfnamefont{G.}~\bibnamefont{Ekström}}, \bibinfo {author}
  {\bibfnamefont{K.}~\bibnamefont{Satake}},\ and\ \bibinfo {author}
  {\bibfnamefont{S.}~\bibnamefont{Sipkin}},\ }%
  \bibfield{journal}{%
  \Doi{10.1126/science.1112250}{\bibinfo {journal} {Science}}\ }%
  \textbf{\bibinfo {volume} {308}},\ \bibinfo {pages} {1127} (\bibinfo {year}
  {2005}),\
  \bibAnnoteFile{NoStop}{lay2005}%
\bibitem{hu2011}%
  \BibitemOpen
  \bibfield{author}{%
  \bibinfo {author} {\bibfnamefont{X.}~\bibnamefont{Hu}}, \bibinfo {author}
  {\bibfnamefont{C.~T.}\ \bibnamefont{Chan}}, \bibinfo {author}
  {\bibfnamefont{K.-M.}\ \bibnamefont{Ho}},\ and\ \bibinfo {author}
  {\bibfnamefont{J.}~\bibnamefont{Zi}},\ }%
  \bibfield{journal}{%
  \Doi{10.1103/PhysRevLett.106.174501}{\bibinfo {journal} {Phys. Rev. Lett.}}\
  }%
  \textbf{\bibinfo {volume} {106}},\ \bibinfo {pages} {174501} (\bibinfo {year} {2011})%
  \bibAnnoteFile{NoStop}{hu2011}%
\bibitem{cavitation}%
  \BibitemOpen
  \bibfield{author}{%
  \bibinfo {author} {\bibfnamefont{J.-P.}\ \bibnamefont{Franc}}\ and\ \bibinfo
  {author} {\bibfnamefont{J.-M.}\ \bibnamefont{Michel}},\ }%
  \Doi{10.1007/1-4020-2233-6}{\emph{\bibinfo {title} {Fundamentals of
  Cavitation}}},\ \bibinfo {series} {Fluid Mechanics and its Applications},
  Vol.~\bibinfo {volume} {76}\ (\bibinfo {publisher} {Kluwer academic
  publishers, Dordrecht},\ \bibinfo {year} {2005})%
  \bibAnnoteFile{NoStop}{cavitation}%
\bibitem{acousticbubble}%
  \BibitemOpen
  \bibfield{author}{%
  \bibinfo {author} {\bibfnamefont{T.}~\bibnamefont{Leighton}},\ }%
  \emph{\bibinfo {title} {The Acoustic Bubble}}\ (\bibinfo {publisher}
  {Academic Press},\ \bibinfo {year} {1994})\ ISBN \bibinfo {isbn}
  {9780124419209},\ 
  \bibAnnoteFile{NoStop}{acousticbubble}%
\bibitem{minnaert1933}%
  \BibitemOpen
  \bibfield{author}{%
  \bibinfo {author} {\bibfnamefont{M.}~\bibnamefont{Minnaert}},\ }%
  \bibfield{journal}{%
  \Doi{10.1080/14786443309462277}{\bibinfo {journal} {Phil. Mag.
  Series 7}}\ }%
  \textbf{\bibinfo {volume} {16}},\ \bibinfo {pages} {235} (\bibinfo {year}
  {1933}),\
  \bibAnnoteFile{NoStop}{minnaert1933}%
\bibitem{kafesaki2000}%
  \BibitemOpen
  \bibfield{author}{%
  \bibinfo {author} {\bibfnamefont{M.}~\bibnamefont{Kafesaki}}, \bibinfo
  {author} {\bibfnamefont{R.~S.}\ \bibnamefont{Penciu}},\ and\ \bibinfo
  {author} {\bibfnamefont{E.~N.}\ \bibnamefont{Economou}},\ }%
  \bibfield{journal}{%
  \Doi{10.1103/PhysRevLett.84.6050}{\bibinfo {journal} {Phys. Rev. Lett.}}\ }%
  \textbf{\bibinfo {volume} {84}},\ \bibinfo {pages} {6050} (\bibinfo {year} {2000}),\
  \bibAnnoteFile{NoStop}{kafesaki2000}%
\bibitem{prosperetti1988}%
  \BibitemOpen
  \bibfield{author}{%
  \bibinfo {author} {\bibfnamefont{A.}~\bibnamefont{Prosperetti}},\ }%
  \bibfield{journal}{%
  \Doi{10.1121/1.396740}{\bibinfo {journal} {J. Acoust. Soc. Am.}}\ }%
  \textbf{\bibinfo {volume} {84}},\ \bibinfo {pages} {1042} (\bibinfo {year}
  {1988}),\ 
  \bibAnnoteFile{NoStop}{prosperetti1988}%
\bibitem{lu1990}%
  \BibitemOpen
  \bibfield{author}{%
  \bibinfo {author} {\bibfnamefont{N.}~\bibnamefont{Lu}}, \bibinfo {author}
  {\bibfnamefont{A.}~\bibnamefont{Prosperetti}},\ and\ \bibinfo {author}
  {\bibfnamefont{S.}~\bibnamefont{Yoon}},\ }%
  \bibfield{journal}{%
  \Doi{10.1109/48.103521}{\bibinfo {journal} {Oceanic Engineering, IEEE Journal
  of}}\ }%
  \textbf{\bibinfo {volume} {15}},\ \bibinfo {pages} {275 } (\bibinfo {year} {1990}).
  \bibAnnoteFile{NoStop}{lu1990}%
\bibitem{prosperetti1977}%
  \BibitemOpen
  \bibfield{author}{%
  \bibinfo {author} {\bibfnamefont{A.}~\bibnamefont{Prosperetti}},\ }%
  \bibfield{journal}{%
  \Doi{DOI:10.1121/1.381252}{\bibinfo {journal} {J. Acoust. Soc. Am.}}\ }%
  \textbf{\bibinfo {volume} {61}},\ \bibinfo {pages} {17} (\bibinfo {year}
  {1977}).
  \bibAnnoteFile{NoStop}{prosperetti1977}%
\bibitem{dagostino1988}%
  \BibitemOpen
  \bibfield{author}{%
  \bibinfo {author} {\bibfnamefont{L.}~\bibnamefont{d'Agostino}}\ and\ \bibinfo
  {author} {\bibfnamefont{C.~E.}\ \bibnamefont{Brennen}},\ }%
  \bibfield{journal}{%
  \Doi{10.1121/1.397058}{\bibinfo {journal} {J. Acoust. Soc. Am.}}\ }%
  \textbf{\bibinfo {volume} {84}},\ \bibinfo {pages} {2126} (\bibinfo {year}
  {1988}),\ 
  \bibAnnoteFile{NoStop}{dagostino1988}%
\bibitem{Note1}%
  \BibitemOpen
  \bibinfo {note} {As bubbles and clouds display deep subwavelength resonances,
  only monopolar behavior has been retained in the expansion.}%
  \bibAnnoteFile{Stop}{Note1}%
\bibitem{morsefeshbach}%
  \BibitemOpen
  \bibfield{author}{%
  \bibinfo {author} {\bibfnamefont{P.~M.}\ \bibnamefont{Morse}}\ and\ \bibinfo
  {author} {\bibfnamefont{H.}~\bibnamefont{Feshbach}},\ }%
  \enquote{\bibinfo {title} {Methods of theoretical physics},}\ \ (\bibinfo
  {publisher} {McGRAW-HILL},\ \bibinfo {year} {1953})\ Chap.~\bibinfo {chapter}
  {11}, pp.\ \bibinfo {pages} {1498--1501}%
  \bibAnnoteFile{NoStop}{morsefeshbach}%
\bibitem{dagostino1989}%
  \BibitemOpen
  \bibfield{author}{%
  \bibinfo {author} {\bibfnamefont{L.}~\bibnamefont{D'Agostino}}\ and\ \bibinfo
  {author} {\bibfnamefont{C.~E.}\ \bibnamefont{Brennen}},\ }%
  \bibfield{journal}{%
  \Doi{10.1017/S0022112089000339}{\bibinfo {journal} {J. Fluid Mech.}}\ }%
  \textbf{\bibinfo {volume} {199}},\ \bibinfo {pages} {155} (\bibinfo {year}
  {1989}),\
  \bibAnnoteFile{NoStop}{dagostino1989}%
\bibitem{caflish1985b}%
  \BibitemOpen
  \bibfield{author}{%
  \bibinfo {author} {\bibfnamefont{R.~E.}\ \bibnamefont{Caflisch}}, \bibinfo
  {author} {\bibfnamefont{M.~J.}\ \bibnamefont{Miksis}}, \bibinfo {author}
  {\bibfnamefont{G.~C.}\ \bibnamefont{Papanicolaou}},\ and\ \bibinfo {author}
  {\bibfnamefont{L.}~\bibnamefont{Ting}},\ }%
  \bibfield{journal}{%
  \Doi{10.1017/S0022112085003354}{\bibinfo {journal} {J. Fluid
  Mech.}}\ }%
  \textbf{\bibinfo {volume} {160}},\ \bibinfo {pages} {1} (\bibinfo {year}
  {1985}),\
  \bibAnnoteFile{NoStop}{caflish1985b}%
\bibitem{miksis1989}%
  \BibitemOpen
  \bibfield{author}{%
  \bibinfo {author} {\bibfnamefont{M.~J.}\ \bibnamefont{Miksis}}\ and\ \bibinfo
  {author} {\bibfnamefont{L.}~\bibnamefont{Ting}},\ }%
  \bibfield{journal}{%
  \Doi{10.1121/1.398442}{\bibinfo {journal} {J. Acoust. Soc. Am.}}\ }%
  \textbf{\bibinfo {volume} {86}},\ \bibinfo {pages} {2349} (\bibinfo {year}
  {1989}),\ 
  \bibAnnoteFile{NoStop}{miksis1989}%
\bibitem{zeravcic2011}%
  \BibitemOpen
  \bibfield{author}{%
  \bibinfo {author} {\bibfnamefont{Z.}~\bibnamefont{Zeravcic}}, \bibinfo
  {author} {\bibfnamefont{D.}~\bibnamefont{Lohse}},\ and\ \bibinfo {author}
  {\bibfnamefont{W.}~\bibnamefont{van Saarloos}},\ }%
  \bibfield{journal}{%
  \Doi{10.1017/jfm.2011.153}{\bibinfo {journal} {J. Fluid Mech.}}\
  }%
  \textbf{\bibinfo {volume} {680}},\ \bibinfo {pages} {114} (\bibinfo {year}
  {2011}),\
  \bibAnnoteFile{NoStop}{zeravcic2011}%
\bibitem{tsunamis2009}%
  \BibitemOpen
  \bibfield{author}{%
  \bibinfo {author} {\bibfnamefont{B.}~\bibnamefont{Levin}}\ and\ \bibinfo
  {author} {\bibfnamefont{M.}~\bibnamefont{Nosov}},\ }%
  \emph{\bibinfo {title} {Physics of Tsunamis}}\ (\bibinfo {publisher}
  {Springer Science + Business Media B.V.},\ \bibinfo {year} {2009})\ pp.\
  \bibinfo {pages} {99--152},\ \bibinfo {note} {role of the Compressibility of
  Water and of Non-linear Effects in the Formation of Tsunami Waves}%
  \bibAnnoteFile{NoStop}{tsunamis2009}%
\bibitem{nosov2007}%
  \BibitemOpen
  \bibfield{author}{%
  \bibinfo {author} {\bibfnamefont{M.~A.}\ \bibnamefont{Nosov}}\ and\ \bibinfo
  {author} {\bibfnamefont{S.~V.}\ \bibnamefont{Kolesov}},\ }%
  \bibfield{journal}{%
  \Doi{10.5194/nhess-7-243-2007}{\bibinfo {journal} {Natural Hazards and Earth
  System Science}}\ }%
  \textbf{\bibinfo {volume} {7}},\ \bibinfo {pages} {243} (\bibinfo {year}
  {2007}),\ 
  \bibAnnoteFile{NoStop}{nosov2007}%
\bibitem{Note2}%
  \BibitemOpen
  \bibinfo {note} {When bubbles are at a depth of a few kilometers, due to the
  increased static pressure all the sizes discussed previously for the
  metacloud will increase in about one order of magnitude.}%
  \bibAnnoteFile{Stop}{Note2}%
\bibitem{hammack1973}%
  \BibitemOpen
  \bibfield{author}{%
  \bibinfo {author} {\bibfnamefont{J.~L.}\ \bibnamefont{Hammack}},\ }%
  \bibfield{journal}{%
  \Doi{10.1017/S0022112073000479}{\bibinfo {journal} {J. Fluid Mech.}}\ }%
  \textbf{\bibinfo {volume} {60}},\ \bibinfo {pages} {769} (\bibinfo {year}
  {1973}),\
  \bibAnnoteFile{NoStop}{hammack1973}%
\bibitem{arcas2012}%
  \BibitemOpen
  \bibfield{author}{%
  \bibinfo {author} {\bibfnamefont{D.}~\bibnamefont{Arcas}}\ and\ \bibinfo
  {author} {\bibfnamefont{H.}~\bibnamefont{Segur}},\ }%
  \bibfield{journal}{%
  \Doi{10.1098/rsta.2011.0457}{\bibinfo {journal} {Phil. Trans. R. Soc. A}}\ }%
  \textbf{\bibinfo {volume} {370}},\ \bibinfo {pages} {1505} (\bibinfo {year}
  {2012}),\
  \bibAnnoteFile{NoStop}{arcas2012}%
\bibitem{ganan2001}%
  \BibitemOpen
  \bibfield{author}{%
  \bibinfo {author} {\bibfnamefont{A.~M.}\ \bibnamefont{Ga\~n\'an Calvo}}\ and\
  \bibinfo {author} {\bibfnamefont{J.~M.}\ \bibnamefont{Gordillo}},\ }%
  \bibfield{journal}{%
  \Doi{10.1103/PhysRevLett.87.274501}{\bibinfo {journal} {Phys. Rev. Lett.}}\
  }%
  \textbf{\bibinfo {volume} {87}},\ \bibinfo {pages} {274501} (\bibinfo {month}
  {Dec}\ \bibinfo {year} {2001}),\
  \bibAnnoteFile{NoStop}{ganan2001}%
\end{thebibliography}

%

\end{document}